\documentclass[prx, preprint, superscriptaddress, endfloats*]{revtex4-2}
\usepackage{amsmath,hyperref,cleveref,amssymb,amsfonts,graphicx,multirow,color,bm,textcomp,mathtools}
\usepackage{times}
\usepackage{amsfonts}
\usepackage{mathrsfs,ulem}
\usepackage{graphicx}
\usepackage{dcolumn}
\usepackage{bm}
\usepackage{xcolor}

\newcommand{\bee}{\begin{equation}}
\newcommand{\ee}{\end{equation}}
\newcommand{\bma}{\begin{pmatrix}}
\newcommand{\ema}{\end{pmatrix}}
\newcommand{\balig}{\begin{align}}
\newcommand{\ealig}{\end{align}}

\newcommand{\bk}{\boldsymbol{k}}

\newcommand{\bZ}{\mathbb{Z}}

\def\be{\begin{equation}}       \def\ee{\end{equation}}
\def\bea{\begin{eqnarray}}      \def\eea{\end{eqnarray}}

\begin{document}
\title{Moving Protocol of  Majorana Corner Modes in a Superconducting 2D Weyl Semimetal Heterostructure}

\author{Ching-Kai Chiu}\email{ching-kai.chiu@riken.jp}
\affiliation {RIKEN Interdisciplinary Theoretical and Mathematical Sciences (iTHEMS), Wako, Saitama 351-0198, Japan}
\author{Yueh-Ting Yao}
\affiliation {Department of Physics, National Cheng Kung University, Tainan, 701, Taiwan}
\author{Tay-Rong Chang}\email{u32trc00@phys.ncku.edu.tw}
\affiliation {Department of Physics, National Cheng Kung University, Tainan, 701, Taiwan}
\affiliation {Center for Quantum Frontiers of Research and Technology (QFort), Tainan, 70101, Taiwan}
\affiliation {Physics Division, National Center for Theoretical Sciences, Taipei, 10617, Taiwan}
\author{Guang~Bian}\email{ biang@missouri.edu}
\affiliation {Department of Physics and Astronomy, University of Missouri, Columbia, Missouri 65211, USA}
\affiliation {MU Materials Science \& Engineering Institute, University of Missouri, Columbia,MO 65211, USA}

\begin{abstract}

Second-order topological superconductors host Majorana corner modes (MCMs), which are confined to specific corners of the system. This spatial restriction presents challenges for manipulating and relocating MCMs. We propose a novel protocol for dynamically controlling the movement of time-reversal symmetric MCMs in a heterostructure consisting of a 2D Weyl semimetal and a $d$-wave superconductor. By leveraging the energy asymmetry of topological edge states in the 2D Weyl semimetal, the position of MCMs can be effectively tuned via chemical potential adjustments. We further introduce a device architecture that integrates multiple heterostructure blocks, each controlled by independent gate voltages, to enable the adiabatic movement and exchange of MCMs. This approach demonstrates a robust mechanism for Majorana manipulation and provides a scalable framework for future experimental studies of topological quantum computation.
\end{abstract}

\maketitle
\newpage

    

\section{Introduction}

The experimental confirmation of Majorana zero modes has advanced gradually, since the initial observation of zero-bias conductance peaks in nanowires~\cite{Mourik_zero_bias,Kitaev2001}. Demonstrating non-Abelian braiding~\cite{Sarma:2015aa,10.21468/SciPostPhysLectNotes.15} of Majorana zero modes is widely regarded as a crucial step toward conclusively verifying their existence. Despite recent progress~\cite{Wang333,2018arXiv181208995M,PhysRevX.8.041056,Li:2022ud,Liu:2024aa,Schneider:2022aa,Uday:2024aa}, several challenges hinder the realization and manipulation of Majorana modes: (1) in-gap Yu-Shiba-Rusinov states~\cite{Yu_magnetic,Shiba:1968vm,Rusinov_magnetic} and Caroli–de Gennes–Matricon modes~\cite{CAROLI1964307} can mimic Majorana signals, complicating experimental verification; and (2) precise manipulation of Majorana modes remains technically demanding. 

On the other hand, Majorana zero modes are highly promising candidates for topological quantum computation due to their inherent topological robustness. Quantum information encoded in spatially separated Majorana modes is topologically protected against local perturbations, providing resilience to decoherence. Furthermore, certain quantum gates can be implemented by braiding Majorana zero modes, leveraging this topological protection. However, universal quantum computation~\cite{DiVincenzo:2000aa,Kitaeve_Univeral} requires additional operations, such as the T-gate, which cannot be realized through Majorana braiding alone. Implementing the T-gate necessitates breaking the topological protection, posing a fundamental trade-off. Recently, ``poor man's" Majorana bound states, which lack full topological protection, have emerged as a potential alternative for manipulating Majorana zero modes~\cite{Zatelli:2024aa,PRXQuantum.5.010323}.

A particularly intriguing platform for realizing Majorana zero modes is second-order topological superconductors (TSCs), which host Majorana zero modes localized at the corners of the system~\cite{PhysRevB.98.245413, PhysRevLett.121.186801}. These second-order TSCs preserving time-reversal symmetry can be engineered by coupling a quantum spin Hall insulator to a proximitized superconductor with $d$-wave pairing. The key mechanism underlying the emergence of Majorana corner modes (MCMs) is the sign change of the superconductivity pairing at the intersection of two adjacent edges, resulting in a time-reversal symmetric Majorana pair with zero energy localized at the pairing domain wall. Despite their theoretical appeal, these MCMs are inherently tied to the specific pairing potential at the corners of the system. This dependence restricts their spatial mobility, posing significant challenges for the effective manipulation and relocation of Majorana zero modes, which are essential for practical applications in scalable topological quantum computing.

In this work, we propose a novel setup for controlling the movement of time-reversal symmetry-protected Majorana pairs in a heterostructure of a 2D Weyl semimetal and a $d$-wave superconductor. Exploiting the asymmetry of the topological edge states in 2D Weyl semimetals, our design enables the topologically protected movement of Majorana pairs by tuning the gate voltage. According to symmetry classifications, time-reversal symmetric Majorana pairs in class DIII are fundamentally distinct from conventional Majorana bound states in class D, which break time-reversal symmetry~\cite{RevModPhys.88.035005}. While the manipulation of individual Majorana bound states has been extensively studied~\cite{PhysRevX.6.031016,Alicea_Majorana_wire,PhysRevResearch.3.023007}, our protocol focuses on the controlled movement and exchange of time-reversal symmetric MCMs, paving the way for new approaches to Majorana manipulation.



\section{2D Weyl semimetal}
\begin{figure}
\centerline{\includegraphics[width=1\textwidth]{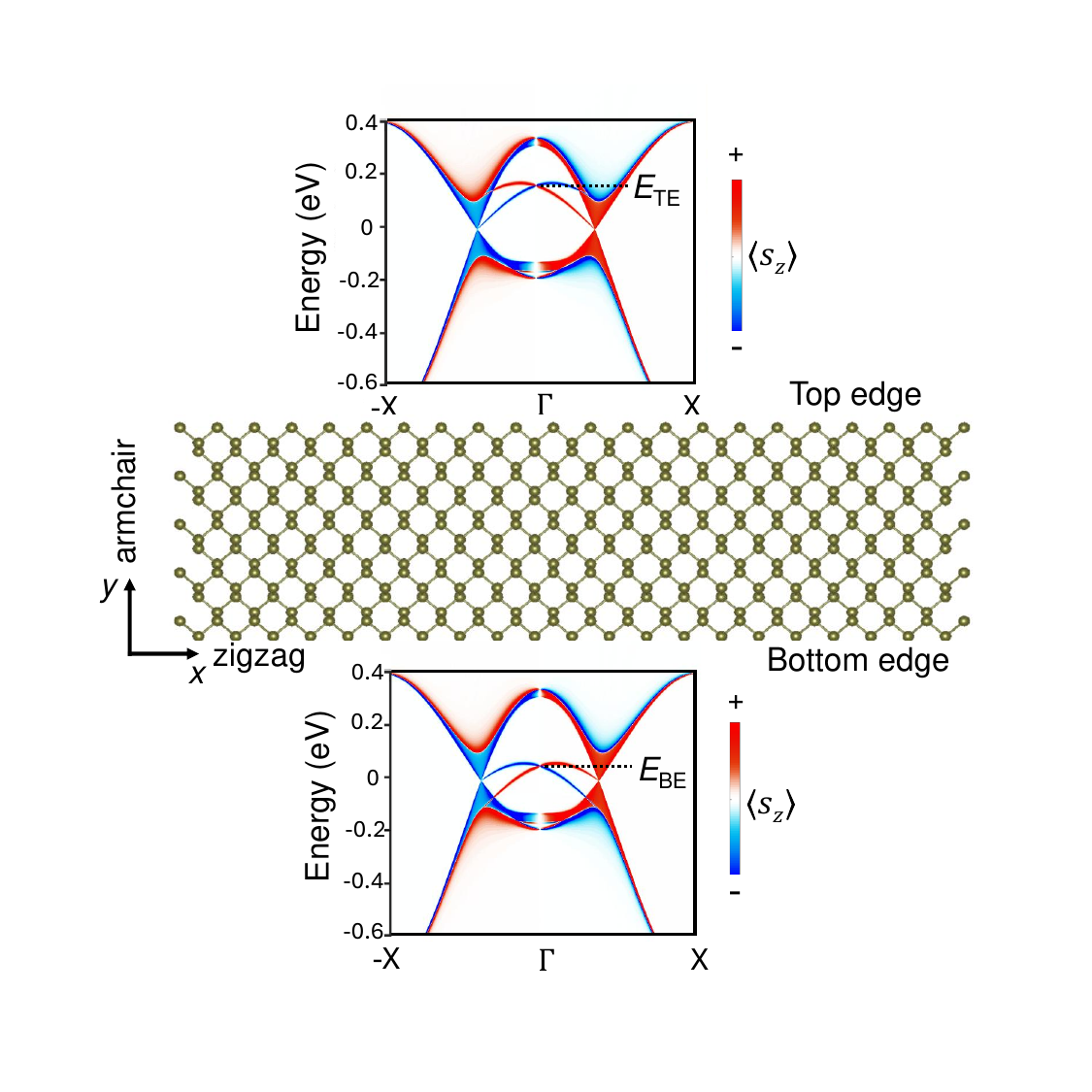}}
\caption{Projected bulk band structure of a 2D Weyl semimetal ribbon along $\mathrm{X}$-$\Gamma$-$\mathrm{X}$ with edge state bands from the top edge (top panel) and the bottom edge (bottom panel). The red and blue colors indicate the spin polarization of $\langle s_z \rangle$. The ball-and-stick diagram at the center illustrates the lattice structure of the 2D Weyl semimetal-epitaxial bismuthene.}
\label{2DWeyl}
\end{figure}

A 2D Weyl semimetal, akin to a spin-polarized variant of graphene, represents a topological matter characterized by Weyl fermion-like quasiparticles in 2D space. The spinful linear band structure in two dimensions gives rise to distinctive topological properties, accompanied by the emergence of Fermi string edge states \cite{PhysRevB.105.075403}. The topological charge of 2D Weyl semimetals can be defined as the winding number of $\pi$, which can be obtained by integrating the Berry phase along a loop encircling each Weyl node \cite{Lu2016}. The nonzero winding number guarantees the existence of topologically protected edge states \cite{Bian2016}. These topological edge states take the form of Fermi strings with one end attached to the projection of bulk Weyl nodes at the Fermi level as shown in Fig.~\ref{2DWeyl}. The highly unusual properties associated with 2D Weyl fermion states have inspired a myriad of theoretical and experimental works\cite{PhysRevLett.118.156401, PhysRevResearch.4.043183, Panigrahi2022, PhysRevB.105.075403, PhysRevB.103.L201115, PhysRevB.106.125404, Mogi2022, adfm202305179, PhysRevLett.131.046601, PhysRevB.107.035122, PhysRevB.108.075160}. An ideal 2D Weyl semimetal with point Fermi surfaces and Fermi string edge states has been observed in the bismuthene monolayer epitaxially grown on SnS(Se) substrates \cite{Lu:2024aa}. The tight-binding Hamiltonian of epitaxial bismuthene can be written as follows:
\begin{align}
    \hat{H}_{\rm 2DWeyl}=&\sum_{\bk} \Big [ C^\dagger_{\bk} \big (\epsilon(\bk) s_0 \tau_0 + c_{1} \sin(bk_y)s_0\tau_y +[c_{2} \cos(ak_x)+c_{3} \cos(bk_y)]s_0\tau_x  \nonumber \\
    &+ \Delta s_0\tau_x + c_{4}\sin(ak_x) s_z\tau_z + \lambda_{\rm D} s_0\tau_z \big )C_{ \bk}   \Big],
  \end{align}
 \noindent where $C_{\bk} = (A_{\uparrow \bk}, A_{\downarrow \bk}, B_{\uparrow \bk}, B_{\downarrow \bk})$ denotes the annihilation operators for the orbitals at A and B sites of bismuthene lattice \cite{Lu:2024aa},  $s_i$ and $\tau_i$ represent the Pauli matrices for the spin and orbital degree of freedom, respectively, the lattice constants of bismuthene (a, b) are (4.5 \AA, 4.8 \AA), and $\epsilon(\bk)=\epsilon_0 +\epsilon_1 \cos(ak_x) +\epsilon_2 \cos(bk_y)$ denote the momentum dependent onsite energy. The parameters (in the unit of eV) $\epsilon_0=1.0749$, $\epsilon_1=0.0585$, $\epsilon_2=-0.0546$, $c_1 = -0.0894$, $c_2 = 0.0939$, $c_3 = 0.2350$, $\Delta = -0.5434$, $c_4 = -0.0256$, $\lambda_{\rm D} = -1.1242$ are determined by fitting to the bulk band structure of bismuthene; the term $\Delta$ controls the band inversion, resulting in two nodal points of bulk bands at ($\pm k_0$, 0) when $\Delta+c_{3}<0$, where $k_0= \frac{1}{a}\cos^{-1}(\frac{-\Delta-c_{3}}{c_{2}})$.  $\lambda_{\rm SOC}\equiv c_{4}\sin({ak_0})$  is defined as the effective spin-orbit coupling, which induces a quantum spin Hall insulating band gap. The term $\lambda_{\rm D}$ is the substrate surface potential, which breaks the inversion symmetry and induces spin splitting in the bulk bands of bismuthene. The system is a gapless 2D Weyl semimetal when $\lambda_{\rm D}=\lambda_{\rm SOC}$. The band structure of a bismuthene ribbon along the zigzag direction is shown in Fig.~\ref{2DWeyl}. The Fermi string edge bands located on two opposite edges are plotted 
separately. The red and blue colors represent spin-up and spin-down polarizations, respectively.  Notably, the Fermi string edge bands disperse differently on the two edges. At the bottom edge, the edge bands connect to the Weyl nodes of the bulk bands at one end of the string and merge into the bulk valence band at the other end, whereas the edge bands at the top edge merge into the bulk conduction band. This connection pattern of edge bands is required by the charge conservation and the nodal bulk band structure.  As discussed in the following section, it is this asymmetry of Fermi string edge bands on opposite edges that enables effective manipulation of Majarana corner modes in the heterostructure device. 

The 2D Weyl semimetal described by Eq.~1 represents a critical state at the transition between two topologically distinct phases, namely, a trivial insulator and a quantum spin Hall insulator. We have the trivial insulator and quantum spin Hall insulator when $\lambda_{\rm D}>\lambda_{\rm SOC}$ and $\lambda_{\rm D}<\lambda_{\rm SOC}$, respectively. This criticality property of 2D Weyl semimetal is schematically illustrated by bulk and edge bands from a simplified tight-binding model in Fig.~\ref{spectrum}. In the trivial insulator, no topological edge state appears within the bulk band gap. By contrast, the edge bands traverse the bulk band gap and connect the conduction and valence bands in the quantum spin Hall insulator. In the 2D Weyl semimetal, the Fermi string edge bands are always attached to the Weyl nodes of the bulk bands at the Fermi level as required by the band topology. We note that the crossing points of the edge bands on opposite edges mismatch in energy as marked by $E_{\rm TE}$ and $E_{\rm BE}$ ($E_{\rm TE}$ = 0.15~eV and $E_{\rm BE}$ = 0.035~eV are the energy of the edge band crossing points as shown in Fig.~\ref{2DWeyl}).

\section{Heterostructure of a $d$-wave superconductor and a 2D Weyl semimetal }

In the literature, the heterostructure of a $d$-wave superconductor and a time-reversal symmetric quantum spin Hall insulator, when configured as a second-order topological superconductor, can support a pair of time-reversal symmetric MCMs at each corner, provided the sample is in a rectangular geometry~\cite{PhysRevLett.121.096803,PhysRevB.98.245413}. The key is that the quantum spin Hall insulator hosts helical electron modes along its edges. MCMs emerge at the domain walls where the $d$-wave superconductor pairing undergoes a sign change, serving as a defect within the helical electron edge modes.

\begin{figure}
\centerline{\includegraphics[width=0.95\textwidth]{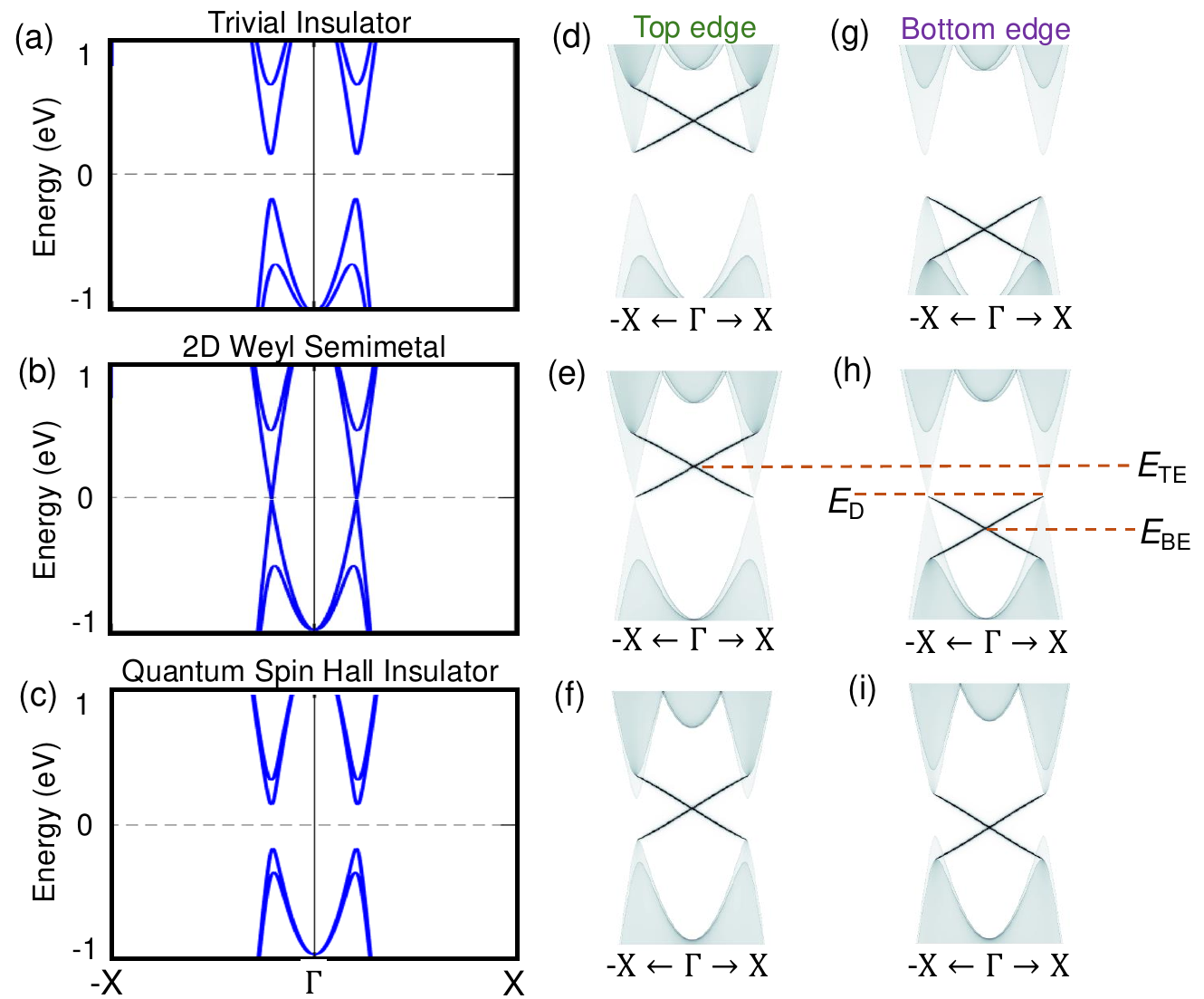}}
\caption{The bulk band structure of (a) a 2D trivial insulator, (b) a 2D Weyl semimetal, and (c) a quantum spin Hall insulator. The projected bulk band structure of (d) a 2D trivial insulator ribbon, (e) a 2D Weyl semimetal ribbon, and (f) a quantum spin Hall insulator ribbon with edge state bands from the top edge. (g-i) Same as (d-f) but with edge state bands from the bottom edge of the ribbon. The energy of the edge band crossing in (e) and (h) is labeled by $E_\mathrm{TE}$ and $E_\mathrm{BE}$, respectively. The energy $E_{\rm D}$ lies between $E_\mathrm{TE}$ and $E_\mathrm{BE}$, marking the point at which the Majorana mode transition to a delocalized state.}
\label{spectrum}
\end{figure}

In contrast, the behavior of a 2D Weyl semimetal~\cite{Lu:2024aa} differs subtly from that of the quantum spin Hall insulator. Due to the broken inversion symmetry, the helical electron modes on opposite edges of the Weyl semimetal reside at different energy levels, as shown in Fig.\ref{spectrum}.b. Consequently, MCMs may not simultaneously appear on opposite edges. This property is crucial, as it enables the adiabatic manipulation of Majorana movement by adjusting the gate voltage within the heterostructure.

\begin{figure}
\centerline{\includegraphics[width=0.95\textwidth]{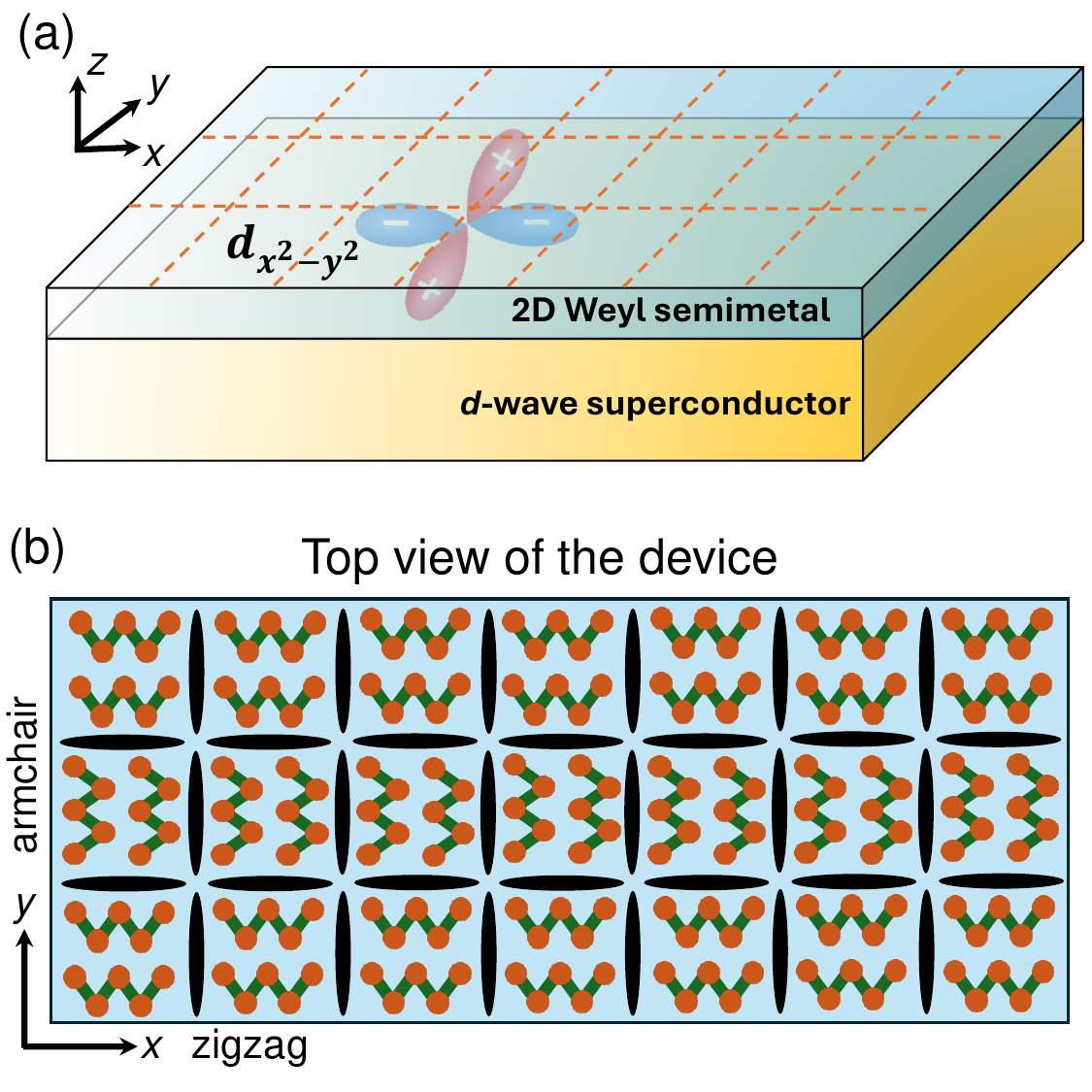}}
\caption{(a) Schematic of a MCM platform that combines a 2D Weyl semimetal with a $d$-wave superconductor. (b) Top view of the MCM device with a grid of 2D Weyl semimetal blocks.  The ball-and-stick diagrams indicate the orientation of the 2D Weyl semimetal blocks. The zigzag direction of the blocks in the top and bottom rows is in the $x$ direction, while the zigzag direction in the middle row is in the $y$ direction. The edges of each square block are weakly coupled with the edges of the neighbor blocks.} 
\label{heterostructure}
\end{figure}

We begin by considering the core physics of a 2D Weyl semimetal with induced $d$-wave superconducting pairing via the proximity effect as illustrated in Fig.\ref{heterostructure}. We note that in the literature the observation of the $d$-wave superconductivity proximity effect is still inconclusive~\cite{Wang:2013aa,PhysRevLett.113.067003}, due to the short coherence lengths of the $d$-wave superconductors. However, recently fractional Shapiro step and time-reversal symmetry breaking have been observed in the Josephson junction of twisted  $d$-wave superconductor layers~\cite{doi:10.1126/science.abl8371}. The Josephson effect hints the emergence of the $d-$wave proximity effect; hence, it is reasonable to introduce the induced $d_{x^2-y^2}$-wave superconducting pairing in the 2D Weyl monolayer.  To model this system, we construct an effective Hamiltonian that captures the superconducting helical edge modes within the heterostructure. The superconducting 2D Weyl Hamiltonian in the form of the second quantization in the momentum space is as follows:
\begin{align}
    \hat{H}_{\rm d}=&\hat{H}_{\rm 2D Weyl}+ \sum_{\bk} \Big [ -\sum_{\sigma=\uparrow,\downarrow}\mu C^\dagger_{ \bk} s_0 \tau_0 C_{ \bk}  \nonumber \\
    &+ \big ( \Delta_d (\cos (a k_x) - \cos (b k_y) ) C^\dagger_{ \bk} is_y \tau_0
    C^\dagger_{ -\bk}   + h.c \big) /2 \Big],
  \end{align}
\noindent where $\hat{H}_{\rm 2D Weyl}$ represents the Hamiltonian for the 2D Weyl semimetal (from Eq.~1). To facilitate the localization of MCMs on a smaller spatial scale, we assign a large value to the $d$-wave gap, $\Delta_d = 0.1$ eV. To solve the Hamiltonian, we rewrite the Hamiltonian in the particle-hole basis $\hat{H}_{\rm d}=\sum_{\bk}(C_{\bk}^\dagger, C_{\bk})H_{\rm BdG}(\bk)(C_{\bk},C_{\bk}^\dagger)^T/2$ as the Bogoliubov-de Gennes formalism and Fourier transform $H_{\rm BdG}(\bk)$ to the real space in the rectangle geometry with four corners as illustrated in Fig.\ref{heterostructure}a. For realistic values of high-temperature superconductors, such as $\Delta_d \sim 0.01$ eV, the core principles of the moving protocol remain applicable.

\begin{figure*}
\centerline{\includegraphics[width=0.95\textwidth]{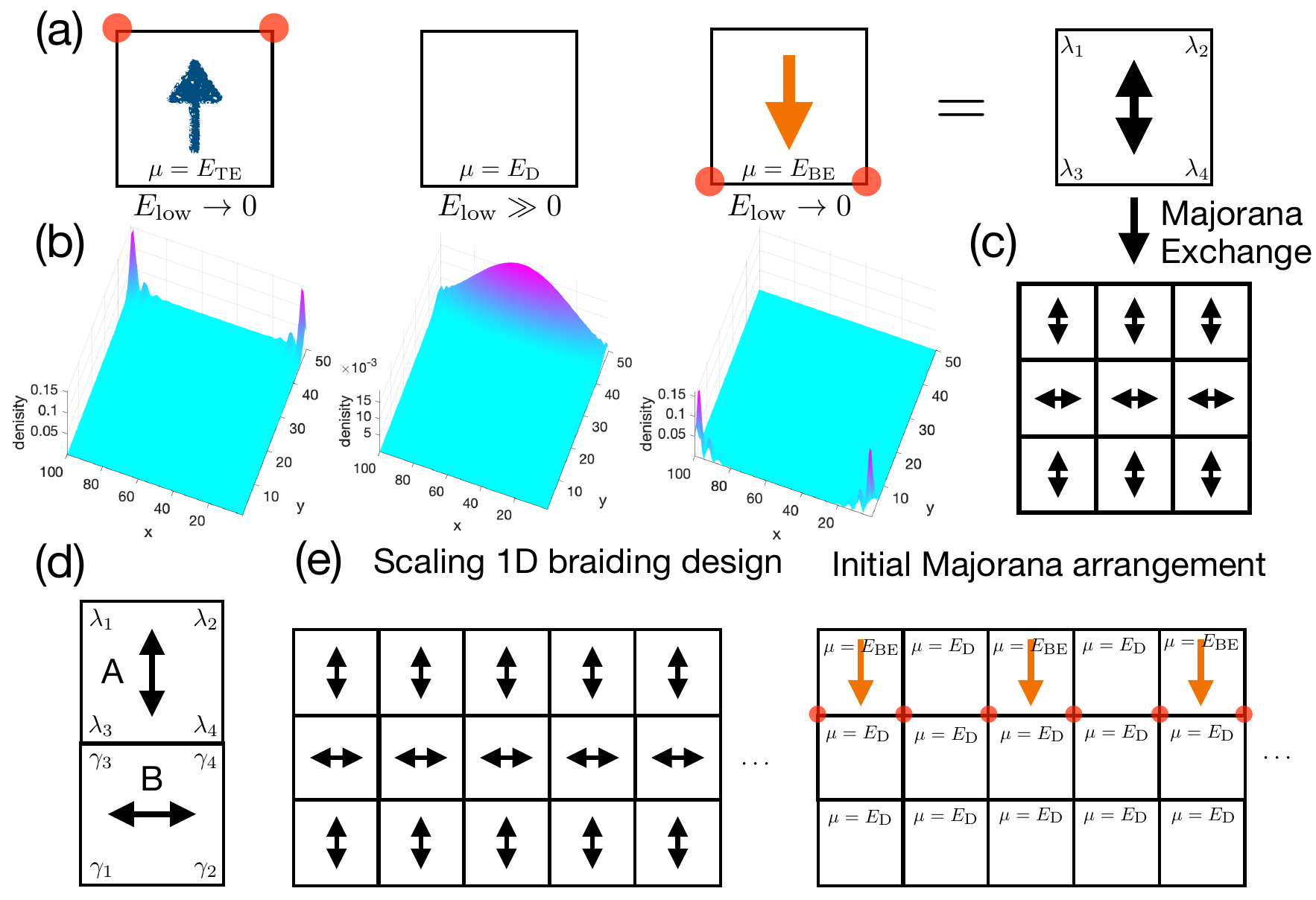}}
\caption{(a) Arrows indicate the corners hosting Majorana pairs ($E_{\rm low}=0$), with the corresponding chemical potential values, while the absence of arrows signifies the vanishing of MCMs. At $\mu=E_{\rm D}$, the lowest energy $E_{\rm low}$ moving away from zero indicates the absence of the Majorana zero modes. The double-headed arrow denotes the orientation of the 2D Weyl semimetal (c.f. Fig.\ref{heterostructure}b). (b) Density distributions of the lowest-energy states corresponding to the three chemical potential values. For $\mu=E_{\rm TE}, E_{\rm BE}$, the corners hosting Majorana modes align with the locations of the helical edge modes. (c) A $3\times 3$ block design is prepared to demonstrate the exchange of the Majorana pairs. (d) A $1\times 2$ block design is to show the Majorana movement that preserves Majorana quantum information. (e) A scalable ${\rm N}\times 3$  block design that facilitates the exchange of any two time-reversal Majorana pairs from the initial arrangement. }
\label{Braiding_setup}
\end{figure*}

Let us go back to normal states without superconductivity first. In the absence of spin mixing terms, $\hat{H}_{\rm 2D Weyl}$ is block-diagonalized by spin up and down, leading to two distinct blocks—(particle up, hole down) and (particle down, hole up), which are connected by time-reversal operation. As illustrated in Fig.\ref{spectrum}, the helical electron modes on the top and bottom edges reside at different energy levels. By setting the chemical potential at $\mu = E_{\rm TE}= 0.15$ eV, we target the helical modes localized on the top edge at $k_x = 0$, while the localized modes on the bottom edge are absent. When we introduce the $d$-wave superconductivity, two pairs of time-reversal symmetric Majorana zero modes (this is evident from the fact that the lowest energy of states in the conduction band $E_{\rm low}=0$) localize solely at the corners of the top edge respectively, as shown in Fig.\ref{Braiding_setup}b.


Next, we adjust the chemical potential to $\mu = E_{\rm D}=0.075$ eV, aligning it with the energy level at which the edge modes are close to the bulk band edge and these modes become notably delocalized on the edge. With 
the $d$-wave superconductivity, this delocalization induces Majorana hybridization, resulting in non-zero energy of the hybridized Majorana modes on the top edge. As shown in Fig.\ref{Braiding_setup}b, the density of these hybridized modes is delocalized. Hence we use the notation \(E_{\rm D}\) to indicate delocalization. To capture the system’s low-energy behavior, we define four Majorana operators associated with the four corners and construct an effective low-energy Hamiltonian to describe the hybridization of these Majorana modes:
\begin{equation}
    \hat{H}_{1 \times 1 } = i A_a(\mu) \lambda_1 \lambda_2 + i A_b(\mu )\lambda_3 \lambda_4.
\end{equation}
Here, the Majorana operators \(\lambda_i\) satisfy \(\lambda_i = \lambda_i^\dagger\) and \(\{\lambda_i, \lambda_j\} = 2\delta_{ij}\). Time-reversal symmetry forces that Majorana operators always appear in pairs. Since the time-reversal partner shares the same spectrum, we consider only one-half of the Hamiltonian for simplicity. Moreover, this effective Hamiltonian shows that when \(\mu = E_{\rm D}\), the coefficients \(A_a(\mu)\) and \(A_b(\mu)\) are nonzero, confirming the absence of Majorana zero modes ($E_{\rm low}\gg 0$).


We further reduce the chemical potential to $\mu =E_{\rm BE}= 0.035$ eV, aligning it with the bottom edge helical modes at $k_x = 0$. At this setting, in the presence of the $d$-wave superconductivity, two pairs of time-reversal symmetric Majorana zero modes ($E_{\rm low}=0$) localize at the two bottom corners. In the effective Hamiltonian, \(A_b(E_{\rm BE}) = 0\), so \(\lambda_3\) and \(\lambda_4\), which commute with \(\hat{H}_{1 \times 1 }\), represent the Majorana zero modes at the two bottom corners, respectively. Similarly, for \(\mu = E_{\rm TE}\), where \(A_a(E_{\rm TE}) = 0\), \(\lambda_1\) and \(\lambda_2\) indicate the Majorana zero modes at the two top corners, respectively.


This enables us to control the presence and location of the MCMs simply by adjusting the gate voltage. In Fig.\ref{Braiding_setup}a, arrows indicate the chemical potential value and the corresponding Majorana locations. It is important to note that this control process does not preserve the quantum information stored in the moving Majoranas within a single block, as the zero-energy modes vanish during the movement. Nevertheless, this setup provides a foundational framework for developing a time-reversal symmetric protocol in the following that enables Majorana manipulation while preserving the encoded quantum information in a multiple-block setup.


To begin, we employ a straightforward design for moving four pairs of MCMs in a heterostructure device, as depicted in Figs.~\ref{heterostructure}b and \ref{Braiding_setup}c, where double-headed arrows indicate the orientation of the heterostructure. The chemical potential $\mu$ of each square block is controlled by a gate independently. Each block of the heterostructure is weakly coupled to its neighboring blocks. The principle behind the movement scheme is to induce hybridization between the two neighboring Majorana wavefunctions, allowing Majorana zero modes to transfer between blocks while remaining zero energy within the gap. In the framework outlined in Eq.~2, the superconductor system can be viewed as two spin-separable, block-diagonal sub-Hamiltonians. Each sub-Hamiltonian, being effectively spinless, belongs to symmetry class BDI. Majorana zero modes in class BDI are strongly protected by the $\mathbb{Z}$ 1D topological invariant, ensuring that within each sub-Hamiltonian, the Majorana energy remains fixed at zero even when multiple Majorana zero modes are located in the same spot. Due to this \(\mathbb{Z}\)-topological invariant protection, Majorana hybridization (or annihilation) is suppressed. Instead of transferring the original MCMs to another block, new Majorana zero modes are created in the target block.


On the other hand, the interface between bismuthene and substrate naturally introduces a Rashba spin-orbit coupling that breaks spin $s_z$ symmetry. Incorporating Rashba spin-orbit coupling transitions the symmetry class of the system from BDI to DIII, enabling the hybridization of Majorana modes from the two spin sub-Hamiltonians. Consequently, the following bulk terms are added to \(\hat{H}_{\rm 2D Weyl}\):
\be
i\lambda_R \sum_{x,y} \big ( C_{x+a,y}^\dagger s_y \tau_z C_{x,y} + C_{x,y+a}^\dagger s_x  \tau_z C_{x,y} \big ),
\ee
where we set the Rashba spin-orbit coupling strength to $\lambda_R = 0.01$ eV in the simulation.
Additionally, at the interfaces between blocks, we assume the coupling between adjacent block edges includes:
\be 
C_{L}^\dagger (e_0 s_0\tau_0 + e_s is_y \tau_z ) C_{R} + iC_{U}^\dagger (e_0 s_0\tau_0 +e_s s_x \tau_z ) C_{O} +h.c.,
\ee
where the first term describes hopping from the right (R) edge to the left (L) edge, and the second term represents hopping from the lower (O) edge to the upper (U) edge. We set the coupling strengths to $e_0 = e_s = 0.1$ eV. In symmetry class DIII, the Majorana pairs are protected only by a \(\mathbb{Z}_2\) topological invariant. As a result, Majorana hybridization induces the movement of Majorana zero modes while preserving the total number of zero modes.

Before demonstrating the Majorana exchange scheme, we first discuss the effective low-energy physics of the Majorana movement that preserves quantum information in a simpler \(1 \times 2\) block arrangement. As illustrated in Fig.~\ref{Braiding_setup}d, block \(B\) is rotated by \(\pi/2\), and the bottom edge of block \(A\) is in contact with the top edge of block \(B\). The operators \(\lambda_i\) describe the MCMs localized at the four corners of block \(A\), while the operators \(\gamma_j\) are for MCMs localized at the four corners of block \(B\). To simplify the analysis, we neglect the time-reversal partners of the Majorana modes. Based on \(\hat{H}_{1\times 1}\), the effective low-energy Hamiltonian as a function of the chemical potentials \((\mu_A, \mu_B)\) in blocks \(A\) and \(B\) can be expressed as:
\begin{align}
\hat{H}_{1\times 2} = & \, i A_a(\mu_A) \lambda_1 \lambda_2 + i A_b(\mu_A) \lambda_3 \lambda_4 + i B_a(\mu_B) \gamma_1 \gamma_3 \nonumber \\
& + i B_b(\mu_B) \gamma_2 \gamma_4 + i d_3(\mu_A, \mu_B) \lambda_3 \gamma_3 + i d_4(\mu_A, \mu_B) \lambda_4 \gamma_4,
\end{align}
where \(B_a(\mu_B)\) and \(B_b(\mu_B)\) exhibit the same chemical-potential dependence as \(A_a(\mu_A)\) and \(A_b(\mu_A)\). Specifically, when \(A_a(E_{\rm TE}) = A_b(E_{\rm BE}) = B_a(E_{\rm TE}) = B_b(E_{\rm BE}) = 0\), MCMs emerge with zero energy at the corresponding edges. The last two terms in the Hamiltonian represent the hybridization of neighboring MCMs. For instance, when \(\mu_A = E_{\rm BE}\) and \(\mu_B = E_{\rm TE}\), \(\lambda_3\) and \(\gamma_3\) are localized near each other, resulting in \(d_3(E_{\rm BE}, E_{\rm TE}) \neq 0\). However, when the chemical potentials \((\mu_A, \mu_B)\) are far from these values, \(d_3\) vanishes due to the delocalization of the Majoranas. Similarly, \(d_4(E_{\rm BE}, E_{\rm BE}) \neq 0\) arises from the overlap of \(\lambda_4\) and \(\gamma_4\), while \(d_4 \sim 0\) elsewhere.

Equipped with this effective Hamiltonian, we can analyze the energy changes during the Majorana movement process, specifically moving a Majorana zero mode from \(\lambda_4\) to \(\gamma_2\). Initially, for \(\mu_A = E_{\rm BE}\) and \(\mu_B = E_{\rm D}\), the effective Hamiltonian
\[
\hat{H}_{1\times 2} = i A_a \lambda_1 \lambda_2 + i B_a \gamma_1 \gamma_3 + i B_b \gamma_2 \gamma_4
\]
shows that \(\lambda_3\) and \(\lambda_4\) are the only Majorana zero modes within the gap, localized at the bottom edge of block \(A\). When we adiabatically adjust \(\mu_B\) from \(E_{\rm D}\) to \(E_{\rm BE}\), the term \(i d_4 \lambda_4 \gamma_4\) appears in the Hamiltonian due to Majorana hybridization, while \(i B_b \gamma_2 \gamma_4\) gradually vanishes. Combining these two terms, \(i(d_4 \lambda_4 + B_b \gamma_2)\gamma_4\), the system evolves such that there are always only two Majorana zero modes: \(\lambda_3\) and \(-\sin \theta \gamma_2 + \cos \theta \lambda_4\), where \(\theta = \tan^{-1}(d_4/B_b)\). As \(\theta\) varies from \(0\) to \(\pi/2\), \(\lambda_4\) adiabatically evolves into \(-\gamma_2\), maintaining zero energy within the gap and preserving the quantum information encoded in the Majoranas.

\begin{figure*}
\centerline{\includegraphics[width=0.95\textwidth]{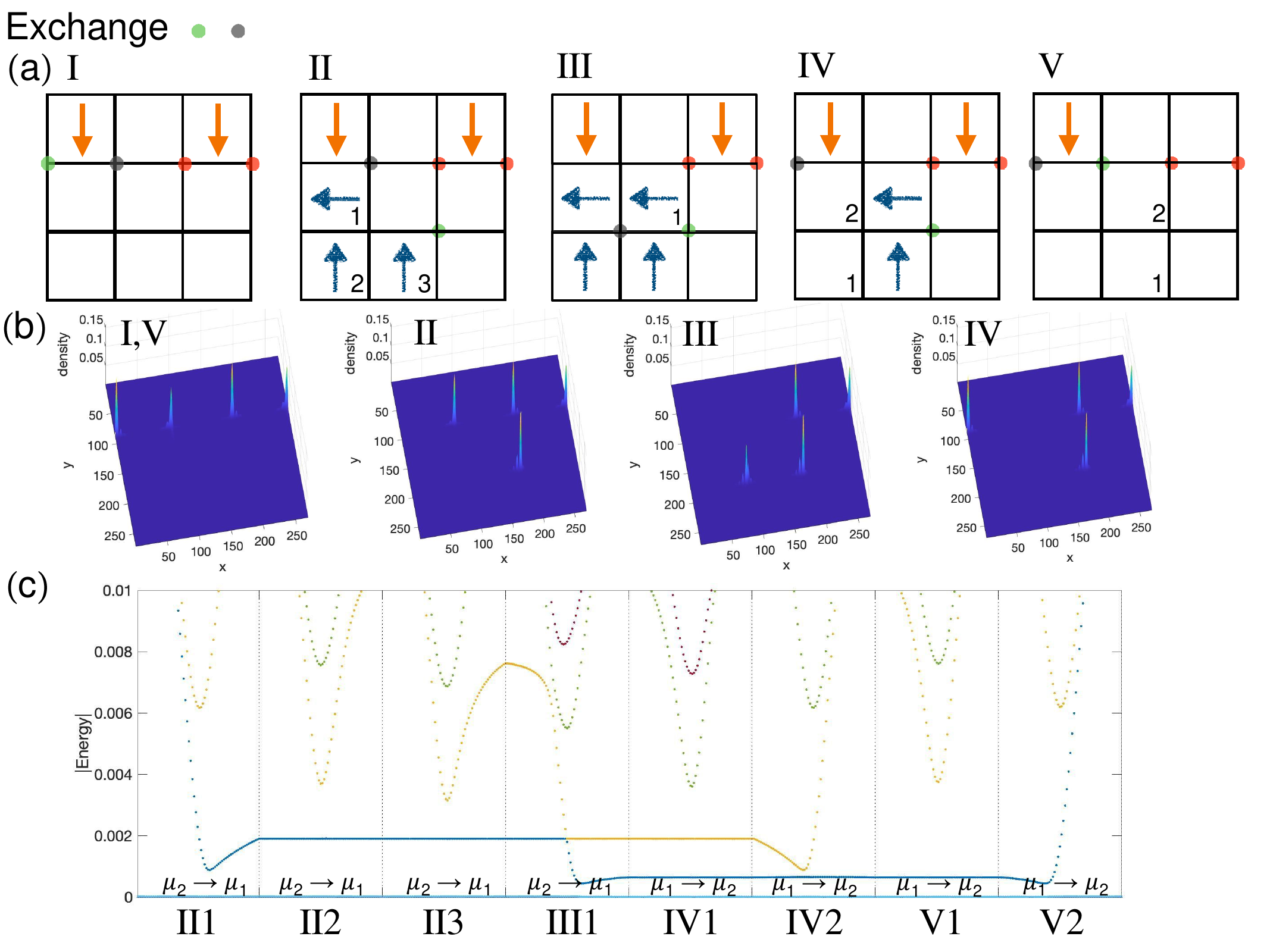}}
\caption{(a) Demonstration of the exchange scheme for the Majorana corner modes depicted in green and gray. Numbers indicate the sequential tuning of chemical potentials, with arrows marking the adjustments within one block at a time. (b) Locations of the Majorana zero modes for the Hamiltonian corresponding to the $3\times 3$ arrangement at each step shown in panel (a). (c) The lowest-energy spectrum throughout the exchange process of tuning the chemical potential. The labels at the bottom indicate the stage in the exchange process as shown in (a). For instance, ``II1" means the block 1 in the step II in (a). The four Majorana pairs remain at zero energy, consistently separated from other states by a persistent energy gap.}
\label{Braiding_check}
\end{figure*}

With this understanding, we can now extend the Majorana exchange scheme to the \(3 \times 3\) block arrangement as illustrated in Fig.\ref{Braiding_setup}c. The exchanging setup comprises three rows of blocks: the first and third rows contain three blocks each, aligned in the original orientation, while each block in the second row is rotated by $\pi/2$. Initially, only two disconnected blocks (with $\mu=E_{\rm BE}$) in the first row host Majorana pairs at their bottom corners, while the remaining blocks are empty (with $\mu=E_{\rm D}$). Following the steps outlined in Fig.\ref{Braiding_check}a-II1, we adiabatically adjust the gate voltage one block at a time. During this process, the initial Majorana pair (for example, the green one) can be annihilated by a newly generated pair in a neighboring block, while a new Majorana pair forms at an alternate corner, which enables the movement of MCMs as discussed in the previous paragraph. By following a similar procedure, we then adjust the gate voltages for blocks II2 and II3 in sequence. Panel b-II shows the positions of the Majorana pairs after adjusting the voltages of the three blocks, which align with the Majorana movement scheme in Fig.\ref{Braiding_check}a-II.




By following the steps in Fig.\ref{Braiding_check}a, the process of Majorana exchange (for the gray and green MCMs) is completed. Throughout this procedure, the MCMs remain at zero energy within the gap, as shown in the simulation in Fig.\ref{Braiding_check}c, allowing the initial Majorana pair to move adiabatically to other blocks as the gate voltage changes. Consequently, the quantum information encoded in the Majorana modes remains protected within the gap. Following the protocol in Fig.\ref{Braiding_check}a, the two Majorana pairs can be adiabatically exchanged. Using a similar protocol, we can exchange the two nearest Majorana pairs, achieving any exchanging for these four pairs.

This setup can be extended into a 1D scalable braiding platform that hosts $2n$ Majorana pairs by constructing a $(2n-1) \times 3 $ block array and initially, in the first row each alternate block hosts two Majorana pairs at the two bottom corners as shown in Fig.\ref{Braiding_setup}e. By adjusting the gate voltage in specific blocks in the second and third rows, it is possible to exchange any two Majorana pairs.

\section{Discussion}

The proposed setup enables the movement of time-reversal symmetric Majorana pairs across short-length scales by simply adjusting gate voltages in different heterostructure blocks. In contrast, existing methods in the literature, such as those relying on spatially varying magnetic fields to move Majorana modes~\cite{PhysRevResearch.3.023007}, encounter significant experimental challenges. Specifically, achieving precise control of magnetic fields with spatially varying directions at the micrometer scale is technically demanding. By comparison, implementing tunable gate voltages for individual blocks is a more practical and experimentally feasible approach. Additionally, our design differs from the well-known Majorana T-junction~\cite{Alicea_Majorana_wire}: while the T-junction involves one-dimensional wires, our heterostructure operates in two dimensions and hosts time-reversal symmetric Majorana pairs localized at block corners.

Although the movement of time-reversal symmetric Majorana pairs does not directly lead to non-Abelian braiding, it represents an important preliminary step toward understanding the Majorana movement features. Two critical characteristics can be experimentally validated: (1) time-reversal symmetry ensures that the energy of an isolated Majorana pair remains pinned at zero, and (2) two Majorana pairs should annihilate each other when brought together.

To verify these properties, scanning tunneling microscopy and transport measurements~\cite{Wang333,Mourik_zero_bias,PhysRevLett.103.237001} can be employed to detect zero-bias conductance peaks, confirming the presence of Majorana pairs. Subsequent application of a magnetic field can break time-reversal symmetry, leading to an energy splitting of the Majorana pairs. Observing this splitting will demonstrate the hybridization of the MCMs, further confirming their topological nature.

Moreover, the proposed setup could be simplified from a $3\times 3$ block arrangement to a 
$1\times 2$ configuration, where the two blocks are oriented at 90 degrees relative to one another. In this simpler setup, Majorana pairs appearing at the common corner of the two blocks should annihilate each other, providing additional evidence of their $\bZ_2$ topological character. In addition, using the concept of poor man's Majorana bound states~\cite{Zatelli:2024aa,PRXQuantum.5.010323}, it is possible to manipulate a physical qubit formed by the splitting of a Majorana pair. 

\section*{Acknowledgments}
C.-K.C. was supported by the Japan Science and Technology Agency (JST) Presto (Grant No. JPMJPR2357), the RIKEN TRIP initiative (RIKEN Quantum), and JST as part of Adopting Sustainable Partnerships for Innovative Research Ecosystem (Grant No. JPMJAP2318). The work at the University of Missouri was supported by the U.S. Department of Energy, Office of Science, Office of Basic Energy Sciences, Division of Materials Science and Engineering, under Grant No. DE-SC0024294. G.B. was supported by the Gordon and Betty Moore Foundation, grant DOI:10.37807/gbmf12247. T.-R.C.was supported by National Science and Technology Council (NSTC) in Taiwan (Program No. MOST111-2628-M-006-003-MY3 and NSTC113-2124-M-006 009-MY3), National Cheng Kung University (NCKU), Taiwan, and National Center for Theoretical Sciences, Taiwan. This research was supported, in part, by the Higher Education Sprout Project, Ministry of Education to the Headquarters of University Advancement at NCKU. T.-R.C. thanks the National Center for High-performance Computing (NCHC) of National Applied Research Laboratories (NARLabs) in Taiwan for providing computational and storage resources.

\bibliography{MCS}
\bibliographystyle{apsrev4-1}

\end{document}